\documentstyle[prb, aps]{revtex}

\draft

\begin{document}

\tighten
\twocolumn[\hsize\textwidth\columnwidth\hsize\csname@twocolumnfalse\endcsname

\title{Aharonov-Bohm effect of excitons in nano-rings}
\author{Hui Hu$^1$, Jia-Lin Zhu$^{1,2}$, Dai-Jun Li$^1$ and Jia-Jiong Xiong$^1$}
\address{$^1$Department of Physics, Tsinghua University, Beijing 100084, P. R. China\\
$^2$Center for Advanced Study, Tsinghua University, Beijing 100084, P. R. China}
\date{\today}
\maketitle

\begin{abstract}
The magnetic field effects on excitons in an InAs nano-ring are studied 
theoretically. By numerically diagonalizing the 
effective-mass Hamiltonian of the problem, which can be separated 
into terms in centre-of-mass and relative coordinates, we calculate the 
low-lying exciton energy levels and oscillator strengths as a function of 
the width of the ring and the strength of the external magnetic field. 
The analytical results are obtained for a narrow-width nano-ring in
which the radial motion is the fastest one and adiabatically decoupled 
from the azimuthal motions. It is shown that in the presence of Coulomb 
correlation, the so called Aharonov-Bohm effect of excitons exists in a 
finite (but small) width nano-ring. However, when the ring width becomes large, 
the non-simply-connected geometry of nano-rings is destroyed and in 
turn yields the suppression of Aharonov-Bohm effect. The conditional 
probability distribution calculated for the low-lying exction states allows 
identification of the presence of Aharonov-Bohm effect.  The linear optical 
susceptibility is also calculated as a function of the magnetic field, to be 
confronted with the future measurements of optical  emission experiments 
on InAs nano-rings. \newline
{PACS numberes: 73.20.Dx, 71.35.-y, 03.65.Bz, 78.66.Fd}\newline
\end{abstract}

]

\section{INTRODUCTION}

Recently, Lorke and collaborators demonstrated the realization of
self-assembled InAs nano-rings inside a completed field-effect transistor 
\cite{lorke98,lorke99,lorke00}. Such small nano-rings (with typical
inner/outer radius of $20/100$ {\rm nm} and $2-3$ {\rm nm} in height) allow
one to study the new non-simply-connected geometry where electrons or holes
could propagate coherently (non-diffusively) all throughout \cite
{mailly,emperador,borrmann,reimann,alfonso,chen,hu,yi,ben,stafford,wend95,wend96a,wend96b,chak,gudm,halonen,niemela}%
. In particular, they offer a unique opportunity to explore the so called
''Aharonov-Bohm effect'' (ABE) of an {\em exciton}, an interesting concept
suggested by Chaplik, R\"{o}mer and Raikh \cite{chaplik,romer}.

Chaplik first predicted the Aharonov-Bohm (AB) oscillation of excitonic
levels in {\em one}-dimensional quantum ring structures \cite{chaplik}, and
most recently, R\"{o}mer and Raikh found the similar results with a short
ranged interaction potential by using a quite different analytical approach 
\cite{romer}. In contrast to the general belief that an exciton, being a
bound state of electron and hole and thus a {\em neutral} entity, should not
be sensitive to the applied flux, they revealed the possibility of a
non-vanishing ABE in one dimensional rings caused by the{\em \ finite}
confinement of an exciton. On the other hand, up to now the influence of the
ring width on ABE has rarely been investigated. Only recently did Song and
Ulloa study the magnetic field effect on excitons in a finite width
nano-ring. They found that the excitons in nano-rings behave to a great
extent as those in quantum dots of similar dimensions and the finite width
of the nano-rings can suppress completely the ABE predicted for
one--dimensional rings \cite{song00}. At the moment, however, to what extent
does the ABE exist in the quasi-one-dimensional (or more important, the {\em %
realistic}) nano-rings is still unclear and is an open subject for research.

In this paper, we would like to investigate systematically the magnetic
field effect on an exciton within a simplified model Hamiltonian, which is
applicable to the {\em realistic} self-assembled semiconducting InAs
nano-rings \cite{lorke98,lorke99,lorke00,pettersson}. By diagonalizing the
effective-mass Hamiltonian of the problem and calculating the low-lying
exciton energy levels and oscillator strengths as a function of the width of
the ring and the strength of the external magnetic field, we show an evident
ABE for nano-rings with a finite but small ring width. The numerical results
are well interpreted with the analytical results for a narrow-width
nano-ring in which the radial motion is the fastest one and adiabatically
decoupled from the azimuthal motions, and also well understood by the
conditional probability distribution calculated for the low-lying exciton
states. The following is a summary of our main results:

(i) In the presence of Coulomb interaction between electrons and holes,
there is no ABE for the ground excitonic state due to the intrinsic
divergence of ground state energy of one-dimensional excitons.

(ii) Some of the low-lying excited energy levels and oscillator strengths of
the exciton show a periodic, Aharonov-Bohm-type oscillation, as a function
of the magnetic field. The periodicity of oscillations is equal to $\Phi
_0\equiv \frac he$ ---the universal flux quantum.

(iii) In addition to the overall blue shifts caused by the diamagnetic
effect, the linear optical susceptibility traces apparently show ABE
oscillations with the magnetic field for some of excited states.

The remainder of this paper is organized as follows. In Sec. II, we present
the simple model for the system and solution method. As a first
approximation, the nano-ring structure is modeled by a double-well like
confinement potential in which we can separate the Hamiltonian into terms in
centre-of-mass and relative coordinates. Subsequently (Sec. III), we give
the analytical results for the narrow-width nano-rings with $m_e^{*}=m_h^{*}$%
. By calculating the tunneling possibility for three typical electron-hole
interaction potential $V_{e-h}(\varphi )$, we explore the underlying physics
of ABE, and the main results of Chaplik, R\"{o}mer and Raikh are recovered.
Sec. IV is devoted to the detailed discussion of numerical results for
different ring widths and magnetic fields, including the low-lying energy
levels, oscillator strengths, conditional probability distributions,
pair-correlation functions, and the linear optical susceptibility. Finally,
we summarize our results in Sec. V.

\section{MODEL\ AND NUMERICAL\ METHOD}

We start from a simplified model Hamiltonian for a two-dimensional exciton
in an InAs nano-ring and in a static magnetic field, simulating recent
experimental nano-ring structures \cite{lorke98,lorke99,lorke00,pettersson}.
The exciton is described by an electron-hole pair ($i=e,h$) with an
effective band edge mass $m_i^{*}$ moving in the x-y plane. The ring-like
structure is well described by a double-well potential, $U\left( {\bf \vec{r}%
}_i\right) =\frac 1{2R_0^2}m_i^{*}\omega _i^2\left( {\bf \vec{r}}%
_i{}^2-R_0^2\right) ^2$, which reproduces a soft barrier $\frac{%
m_i^{*}\omega _i^2R_0^2}2$\ at the center of the sample produced by
self-assembly \cite{lorke98,lorke99,lorke00,pettersson}. Here, $R_0$ is the
radius of the ring and $\omega _i$ is the characteristic frequency of the
radial confinement, giving a characteristic ring width $W\approx 2\sqrt{%
\frac \hbar {2m_i^{*}\omega _i}}$ for each particle. The whole system is
subjected to an external magnetic field perpendicular to the x-y plane. The
resulting model Hamiltonian is thus given by: 
\begin{equation}
{\cal H}=\sum\limits_{i=e,h}\left[ \frac{\left( {\bf \vec{p}}_i+q_i{\bf \vec{%
A}}({\bf \vec{r}}_i)\right) ^2}{2m_i^{*}}+U\left( {\bf \vec{r}}_i\right)
\right] -\frac{e^2}{4\pi \varepsilon _0\varepsilon _r\left| {\bf \vec{r}}_e-%
{\bf \vec{r}}_h\right| },  \label{totHami}
\end{equation}
where ${\bf \vec{r}}_i=\left( x_i,y_i\right) $ and ${\bf \vec{p}}_i=-i\hbar 
{\bf \vec{\nabla}}_i$ denote the position vector and momentum operator, $%
\varepsilon _0$ is the vacuum permittivity, and $\varepsilon _r$ is the
static dielectric constant of the host semiconductor. $q_e=-e$ and $q_h=+e$.
We use symmetric gauge to introduce the external magnetic field, {\em i.e.}, 
${\bf \vec{A}}({\bf \vec{r}}_i)=\frac 12{\bf \vec{B}}\times {\bf \vec{r}}_i$.

It should be pointed out that the present {\it double-well} like confinement
potential can be rewritten as $U\left( {\bf \vec{r}}_i\right) =\frac 12%
m_i^{*}\omega _i^2\left( r_i{}-R_0\right) ^2\frac{\left( r_i{}+R_0\right) ^2%
}{R_0^2}$. If one replaces the operator $r_i$ in factor $\frac{\left(
r_i{}+R_0\right) ^2}{R_0^2}$ by its mean value $\left\langle
r_i\right\rangle =R_0$, the confinement potential returns to the widely used
parabolic form \cite{lorke00,emperador,borrmann,hu,chak,halonen,song00}. On
the other hand{\bf ,} in the limit of the small radius $R_0$ or for a small
potential strength $\omega _i$, the soft barrier at the ring center is very
weak, and the description of the nano-ring is more close to that of a
quantum dot. For a fixed ring width (or potential strength), the crossover
from nano-rings to quantum dots is determined by $R_0\sim \frac{\sqrt{2}}2W$%
\ or $\frac{\hbar \omega _i}2\sim \frac{m_i^{*}\omega _i^2R_0^2}2$, which
means the lowest energy of radial confinement is comparable to the soft
barrier at the ring center{\bf .} It should also be pointed out that our
double-well like confinement potential has been used to calculate the
far-infrared spectroscopy for a two-electron nano-ring \cite{unpublish}, in
good agreement with the recent experiment done by Lorke{\bf \ }{\em et al}%
{\bf . }\cite{lorke00}. In Fig. 1, we show the schematic geometry of InAs
nano-rings with radius $R_0=20$ {\rm nm} for electrons. (a) and (b)
correspond to two different ring widths $W=10$ and $7$ {\rm nm},
respectively.

In terms of the relative coordinate ${\bf \vec{r}}={\bf \vec{r}}_e-{\bf \vec{%
r}}_h$ and center-of-mass coordinate ${\bf \vec{R}}=\frac{m_e^{*}{\bf \vec{r}%
}_e+m_h^{*}{\bf \vec{r}}_h}{m_e^{*}+m_h^{*}}$ the model Hamiltonian can be
separated into the motion of center of mass, relative motion of the
electron-hole pair, and the mixed part: 
\begin{eqnarray}
{\cal H} &=&{\cal H}_{cm}\left( {\bf \vec{R}}\right) +{\cal H}_{rel}\left( 
{\bf \vec{r}}\right) +{\cal H}_{mix}\left( {\bf \vec{R},\vec{r}}\right) 
\nonumber \\
{\cal H}_{cm}\left( {\bf \vec{R}}\right) &=&\frac{{\bf \vec{P}}_{cm}^2}{2M}+%
\frac 12M\omega _{cm}^2\frac{\left( {\bf \vec{R}}{}^2-R_0^2\right) ^2}{R_0^2}
\nonumber \\
{\cal H}_{rel}\left( {\bf \vec{r}}\right) &=&\frac{{\bf \vec{p}}_{rel}^2}{%
2\mu }+\frac{e^2B^2}{8\mu }r^2  \nonumber \\
&&+\frac{eB}2\left( \frac 1{m_e^{*}}-\frac 1{m_h^{*}}\right) {\bf \vec{r}%
\times \vec{p}}_{rel}  \nonumber \\
&&+\frac \mu 2\frac{\left( m_h^{*3}\omega _e^2+m_e^{*3}\omega _h^2\right) }{%
M^3}\frac{r^4}{R_0^2}  \nonumber \\
&&-\mu \omega _{rel}^2{}r^2-\frac{e^2}{4\pi \varepsilon _0\varepsilon _rr} 
\nonumber \\
{\cal H}_{mix}\left( {\bf \vec{R},\vec{r}}\right) &=&\frac{eB}M{\bf \vec{r}%
\times \vec{P}}_{cm}  \nonumber \\
&&-2\mu \left( \omega _e^2-\omega _h^2\right) \left( {\bf \vec{R}\cdot \vec{r%
}-}\frac{{\bf \vec{R}}^3{\bf \cdot \vec{r}}}{R_0^2}\right)  \nonumber \\
&&+\mu \omega _{rel}^2\frac{\left[ R^2r^2+2\left( {\bf \vec{R}\cdot \vec{r}}%
\right) ^2\right] }{R_0^2}  \nonumber \\
&&+2\mu \frac{\left( m_h^{*2}\omega _e^2-m_e^{*2}\omega _h^2\right) }{M^2}%
\frac{{\bf \vec{R}\cdot \vec{r}}^3}{R_0^2},
\end{eqnarray}
where $\mu =\frac{m_e^{*}m_h^{*}}M$ is the electron-hole reduced mass and $%
M=m_e^{*}+m_h^{*}$ is the total mass. We have also introduced a
center-of-mass frequency $\omega _{cm}=\sqrt{\frac{\left( m_e^{*}\omega
_e^2+m_h^{*}\omega _h^2\right) }M}$ and a relative frequency $\omega _{rel}=%
\sqrt{\frac{m_h^{*}\omega _e^2+m_e^{*}\omega _h^2}M}$.

The main purpose in the change of variable above is to use the solutions of $%
H_{cm}$ and $H_{rel}$ as a basis for solving the full Hamiltonian. Those
solutions, {\em i.e.}, labeled by $\psi _\lambda ^{cm}\left( {\bf \vec{R}}%
\right) $ and $\psi _{\lambda ^{\prime }}^{rel}\left( {\bf \vec{r}}\right) $%
, can be solved by the series expansion method \cite{zhu90,zhu97}. Here, $%
\lambda =\left\{ n_{cm},l_{cm}\right\} $ and $\lambda ^{\prime }=\left\{
n_{rel},l_{rel}\right\} $ represent the quantum number pair of the radial
quantum number $n$ and orbital angular-momentum quantum number $l$. Another
advantage of the separation of center-of-mass and relative coordinates is
that the negative Coulomb interaction $-\frac{e^2}{4\pi \varepsilon
_0\varepsilon _rr}$\ appears in{\bf \ }$H_{rel}$ only{\bf , }and the
well-known poor-convergence of the parabolic basis is thus avoided when the
characteristic scale of systems is beyond the effective Bohr radius{\bf \ }%
\cite{song00,song95}. We now search for the wave functions of the exciton in
the form 
\begin{equation}
\Psi =\sum\limits_{\lambda ,\lambda ^{\prime }}A_{\lambda ,\lambda ^{\prime
}}\psi _\lambda ^{cm}\left( {\bf \vec{R}}\right) \psi _{\lambda ^{\prime
}}^{rel}\left( {\bf \vec{r}}\right) .  \label{numwvf}
\end{equation}
Due to the cylindrical symmetry of the problem, the exciton wave functions
can be labeled by the total orbital angular momentum $L=l_{cm}+l_{rel}$. To
obtain the coefficients $A_{\lambda ,\lambda ^{\prime }}$, the total
Hamiltonian is diagonalized in the space spanned by the product states $\psi
_\lambda ^{cm}\left( {\bf \vec{R}}\right) \psi _{\lambda ^{\prime
}}^{rel}\left( {\bf \vec{r}}\right) $. In the present calculations, we first
solve the single particle problem of center-of-mass and relative
Hamiltonians $H_{cm}$ and $H_{rel}$, keep several hundreds of the single
particle states, and then pick up the low-lying energy levels to construct
several thousands of product states. Note that our numerical diagonalization
scheme is very efficient and essentially exact in the sense that the
accuracy can be improved as required by increasing the total number of
selected product states.

Once the coefficients $A_{\lambda ,\lambda ^{\prime }}$ are obtained, one
can calculate directly the measurable properties, such as the linear optical
susceptibility of the nano-rings, whose imaginary part is related to the
absorption intensity measured by optical emission experiments. In theory,
the linear optical susceptibility is proportional to the dipole matrix
elements between one electron-hole pair $j$ state and the vacuum state,
which in turn is proportional to the oscillator strengths $F_j$. In the
dipole approximation, it is given by \cite{song95,bryant,que} 
\begin{eqnarray}
F_j &=&\left| \int \int d{\bf \vec{R}}d{\bf \vec{r}}\Psi \left( {\bf \vec{R},%
\vec{r}}\right) \delta \left( {\bf \vec{r}}\right) \right| ^2  \nonumber \\
&=&\left| \sum\limits_{\lambda ,\lambda ^{\prime }}A_{\lambda ,\lambda
^{\prime }}\psi _{\lambda ^{\prime }}^{rel}\left( {\bf 0}\right) \int d{\bf 
\vec{R}}\psi _\lambda ^{cm}\left( {\bf \vec{R}}\right) \right| ^2,
\end{eqnarray}
where the factor $\psi _{\lambda ^{\prime }}^{rel}\left( {\bf 0}\right) $
and the integral over ${\bf \vec{R}}$ ensure that only the excitons with $%
L=0 $ are created by absorbing photons. Therefore, the frequency dependence
of the linear optical susceptibility $\chi \left( \omega \right) $ can be
expressed as \cite{song95,bryant,que} 
\begin{equation}
\chi \left( \omega \right) \propto \sum_j\frac{F_j}{\hbar \omega
-E_g-E_j-i\Gamma },
\end{equation}
where $E_g$ and $E_j$ are the respective semiconducting band gap of InAs and
energy levels of the exciton, and $\Gamma $ has been introduced as a
phenomenological broadening parameter.

\section{NARROW WIDTH NANO-RINGS}

In order to explore the underlying physics of ABE, we first consider a
narrow-width nano-ring in which the analytical results are variable. We
restrict ourselves with the condition $m_e^{*}=m_h^{*}=m^{*}$, but the
general properties of nano-rings are not affected by this constraint. The
technique employed in this section follows directly the previous works of
Wendler, Fomin and Chaplik describing the rotating Wigner molecule behavior
in quantum rings \cite{wend95,wend96a,wend96b,note}.

By introducing polar coordinates in the x-y plane ${\bf \vec{r}}%
_i=(r_i,\varphi _i)$, the relative azimuthal coordinate $\varphi =\varphi
_e-\varphi _h$, and the azimuthal coordinate $\Theta =\frac{\varphi
_e+\varphi _h}2$ describing the motion of the electron-hole pair as a whole,
the Hamiltonian (\ref{totHami}) reads 
\begin{eqnarray}
{\cal H} &=&\sum\limits_{i=e,h}\left\{ \left[ -\frac{\hbar ^2}{2m^{*}}\left( 
\frac{\partial ^2}{\partial r_i^2}+\frac 1{r_i}\frac \partial {\partial r_i}%
\right) +U\left( r_i\right) \right] \right\}  \nonumber \\
&&-\frac{\hbar ^2}{2m^{*}}\left\{ \frac 1{r_e^2}\left( \frac \partial {%
\partial \varphi }-i\frac{\Phi _e}{\Phi _0}\right) ^2+\frac 1{r_h^2}\left( 
\frac \partial {\partial \varphi }-i\frac{\Phi _h}{\Phi _0}\right) ^2\right.
\nonumber \\
&&+\left. \left( \frac 1{r_e^2}+\frac 1{r_h^2}\right) \frac 14\frac{\partial
^2}{\partial \Theta ^2}+\left( \frac 1{r_e^2}-\frac 1{r_h^2}\right) \frac %
\partial {\partial \varphi }\frac \partial {\partial \Theta }\right\} 
\nonumber \\
&&+V_{e-h}\left( \left| {\bf \vec{r}}_e-{\bf \vec{r}}_h\right| \right) ,
\label{transHami}
\end{eqnarray}
where $\Phi _e=\pi Br_e^2$ and $\Phi _h=\pi Br_h^2$. For a narrow-width
nano-ring, {\em i.e.}, $W\ll R_0$, the radial motion is much faster than the
azimuthal motions. Hence the radial motion is {\em adiabatically decoupled}
from the azimuthal motions with the result 
\begin{equation}
\Psi (r_{1,}r_2;\varphi ,\Theta )=\sum\limits_{n_1,n_2}\Xi
_{n_1,n_2}(r_{1,}r_2)\psi _{n_1,n_2}(\varphi ,\Theta )  \label{totwvf}
\end{equation}
for the excitonic states. $\Xi _{n_1,n_2}(r_{1,}r_2)$ is a product of the
corresponding single-particle wave functions $\chi _{n_i}(r_i)$ with
eigenenergies $\varepsilon _{n_i}^{rad}$ ($n_i=0,1,2,...$), which describe
the radial motion of electrons or holes with zero angular momentum. Because
the single-particle wave functions are orthonormalized, the set $\{\Xi
_{n_1,n_2}(r_{1,}r_2)\}$ forms a closure set of orthonormalized functions.
The azimuthal wave function should satisfy the {\em single-valuedness
boundary conditions}, {\em i.e}, $\psi _{n_1,n_2}(\varphi _1,\varphi
_2)=\psi _{n_1,n_2}(\varphi _1+2\pi ,\varphi _2)=\psi _{n_1,n_2}(\varphi
_1,\varphi _2+2\pi )=\psi _{n_1,n_2}(\varphi _1+2\pi ,\varphi _2+2\pi )$, or
in terms of the new variables $\varphi $ and $\Theta ,$ $\psi
_{n_1,n_2}(\varphi ,\Theta )=\psi _{n_1,n_2}(\varphi +2\pi ,\Theta +\pi
)=\psi _{n_1,n_2}(\varphi ,\Theta +2\pi ).$

As long as the above-stated criterion of the adiabatic approximation is
satisfied, the excited states of radial motion lie high above the ground
state. As a consequence we can restrict the consideration to the ground
state of the radial motion because here only the lowest lying states are of
interest. Thus we take only $n_1=n_2=0$ and substitute Eq. (\ref{totwvf})
into the Schr\"{o}dinger equation ${\cal H}\Psi =E\Psi $ with the
transformed Hamiltonian of Eq. (\ref{transHami}), multiply both parts by $%
\Xi _{0,0}(r_{1,}r_2)$ and integrate over $r_{1,}r_2$. Hence the variables $%
\varphi $ and $\Theta $ become separated: 
\begin{eqnarray}
&&\left\{ -\frac{\hbar ^2}{m^{*}}\left\langle \frac 1{r_1^2}\right\rangle
\left( \left( \frac \partial {\partial \varphi }-i\frac \Phi {\Phi _0}%
\right) ^2+\frac 14\frac{\partial ^2}{\partial \Theta ^2}\right) \right. 
\nonumber \\
&&+\left\langle V_{e-h}\left( \sqrt{r_1^2+r_2^2-2r_1r_2\cos \varphi }\right)
\right\rangle  \nonumber \\
&&+\left. 2\varepsilon _0^{rad}-E\right\} \psi _{0,0}(\varphi ,\Theta )=0,
\label{aziequ}
\end{eqnarray}
where $\Phi =\left\langle \Phi _e\right\rangle =\left\langle \Phi
_h\right\rangle $ in the first order approximation, and $\left\langle
{}\right\rangle $ denotes the average with the radial wave functions 
\begin{equation}
\left\langle ...\right\rangle =\int dr_1r_1\int dr_2r_2\Xi
_{0,0}^{*}(r_{1,}r_2)...\Xi _{0,0}(r_{1,}r_2).
\end{equation}
The magnetic flux in Eq. (\ref{aziequ}) can be removed by a gauge
transformation with the price of introducing the {\em twisted boundary
conditions}: 
\begin{eqnarray}
\psi _{0,0}(\varphi ,\Theta ) &=&e^{i\frac{2\pi \Phi }{\Phi _0}}\psi
_{0,0}(\varphi +2\pi ,\Theta +\pi )  \nonumber \\
\psi _{0,0}(\varphi ,\Theta ) &=&\psi _{0,0}(\varphi ,\Theta +2\pi )
\label{bc}
\end{eqnarray}
It is obvious from Eq. (\ref{aziequ}) that the relative azimuthal motion is
decoupled from the azimuthal motion of the electron-hole pair as a whole.
Therefore the azimuthal wave function can be represented in the form 
\begin{equation}
\psi _{0,0}(\varphi ,\Theta )=\Phi _j^p(\varphi )Q_\nu (\Theta ),
\label{aziwvf}
\end{equation}
where 
\begin{equation}
Q_\nu (\Theta )=\frac 1{\sqrt{2\pi }}e^{i\nu \Theta },  \label{wholewvf}
\end{equation}
and $\Phi _j^p(\varphi )$ is a solution of the equation 
\begin{equation}
\left\{ -\frac{\hbar ^2}{m^{*}R_0^2}\frac{\partial ^2}{\partial \varphi ^2}%
+V_{e-h}\left( 2R_0\left| \sin \frac \varphi 2\right| \right) -\varepsilon
_j^{r.a.,p}\right\} \Phi _j^p(\varphi )=0.
\end{equation}
Here we have used 
\begin{eqnarray*}
\left\langle \frac 1{r_1^2}\right\rangle &=&\frac 1{R_0^2}, \\
\left\langle V_{e-h}\left( \sqrt{r_1^2+r_2^2-2r_1r_2\cos \varphi }\right)
\right\rangle &=&V_{e-h}\left( 2R_0\left| \sin \frac \varphi 2\right|
\right) .
\end{eqnarray*}
The index $p$ describes the possible symmetry types of $\Phi _j^p(\varphi )$
and can be specified below by the twisted boundary condition. In Fig. 2, we
show three typical types of electron-hole interaction potentials $%
V_{e-h}\left( 2R_0\left| \sin \frac \varphi 2\right| \right) $ with $2\pi $
periodicity. In the view of the excitonic state, the potential should be
strong enough to bind the electron-hole pair, and then the relative
azimuthal motion is strongly localized in each potential well. Thus the
relative azimuthal wave function can be considered in the tight-binding-like
form 
\begin{equation}
\Phi _j^p(\varphi )=\sum\limits_{Q=-\infty }^{+\infty }\exp (ip\varphi
_Q)\phi _j(\varphi -\varphi _Q),  \label{rawvf}
\end{equation}
where $\varphi _Q=2\pi Q;Q=0,\pm 1,\pm 2,....$ The wave function $\phi
_j(\varphi -\varphi _Q)$ of a single well satisfies the equation 
\begin{equation}
\left\{ -\frac{\hbar ^2}{m^{*}R_0^2}\frac{\partial ^2}{\partial \varphi ^2}%
+V_{e-h}^{SW,Q}\left( \varphi \right) -\varepsilon _j^{r.a.}\right\} \phi
_j(\varphi -\varphi _Q)=0,
\end{equation}
where $V_{e-h}^{SW,Q}\left( \varphi \right) $ is the potential of a given
single well with minimum at $\varphi =\varphi _Q$ which coincides with the
periodic potential $V_{e-h}\left( 2R_0\left| \sin \frac \varphi 2\right|
\right) $ in the region $\left| \varphi -\varphi _Q\right| <\pi $. Note that
the tight-binding-like relative azimuthal wave function satisfies the
Floquet-Bloch theorem, which is applicable to the periodic potential.

Combining Eqs. (\ref{wholewvf}), (\ref{rawvf}) and (\ref{aziwvf}) and
substituting them into the twist boundary condition Eq. (\ref{bc}), we
obtain 
\begin{eqnarray}
\nu &=&%
\mathop{\rm integer}%
,  \nonumber \\
p+\frac \nu 2+\frac \Phi {\Phi _0} &=&%
\mathop{\rm integer}%
.
\end{eqnarray}
It becomes obvious that $\nu $ describes the total angular momentum of
excitons.

The tunneling through the potential from well to well leads to an explicit
dependence of the energy levels $\varepsilon _j^{r.a.,p}$ on the symmetry
parameter $p$ and thus results in {\em energy bands} as a function of the
magnetic flux $\Phi $. In the tight-binding approximation, it is given by 
\begin{equation}
\varepsilon _j^{r.a.,p}=\varepsilon _j^{r.a.}-t_{0,j}-2t_{1,j}\cos \left(
2\pi \frac \Phi {\Phi _0}+\pi \nu \right) ,  \label{miniband}
\end{equation}
where 
\begin{eqnarray}
t_{0,j} &=&-\int d\varphi \phi _j(\varphi -\varphi _0)\left[ V_{e-h}\left(
\varphi \right) -V_{e-h}^{SW,0}\left( \varphi \right) \right] \phi
_j(\varphi -\varphi _0),  \nonumber \\
t_{1,j} &=&-\int d\varphi \phi _j(\varphi -\varphi _0)\left[ V_{e-h}\left(
\varphi \right) -V_{e-h}^{SW,0}\left( \varphi \right) \right] \phi
_j(\varphi -\varphi _1).  \label{tj}
\end{eqnarray}
It is obvious from Eq. (\ref{miniband}) that the hopping integral $t_{1,j}$
determines the width of the energy band $\varepsilon _j^{r.a.,p}$ and causes
the Aharonov-Bohm-like oscillation in energy levels. However, as follows
from the detailed discussion of this effect below, $t_{1,j}$ is typically
exponentially small compared with $\varepsilon _j^{r.a.,p}$. Only when $%
\varepsilon _j^{r.a.}$ is very close to the maximum of interaction potential 
$V_{e-h}\left( \pi \right) $, does $t_{1,j}$ have a finite value. In other
words, {\em in order to access the measurable AB effect, the bound electron
and hole should have a possibility to tunnel in the opposite directions} 
{\em and meet each other ''on the opposite side of the nano-ring'' (}$%
\varphi =\pi ${\em )}.

Using the wave function (\ref{aziwvf}) of a state with a fixed angular
momentum $\nu $ in the Schr\"{o}dinger equation (\ref{aziequ}), we finally
obtain the eigenenergies of the full Hamiltonian, 
\begin{equation}
E_{0,0,p,j,\nu }=2\varepsilon _0^{rad}+\varepsilon _j^{r.a.,p}+\frac{\hbar
^2\nu ^2}{4m^{*}R_0^2},  \label{spectra}
\end{equation}
and the expression for the oscillation strengths 
\begin{equation}
F_j=\frac{\left| \Phi _j^p(0)\right| ^2}{\int\limits_{-\pi }^{+\pi }d\varphi
\left| \Phi _j^p(\varphi )\right| ^2}\delta _{\nu ,0}.  \label{osc}
\end{equation}
In the following we consider the detailed examples for three typical
electron-hole interaction potentials, restricting ourselves with $\nu =0$.

\subsection{short ranged potential}

We use the $\delta $ potential to simulate the short ranged potential, {\em %
i.e.} $V_{e-h}\left( 2R_0\left| \sin \frac \varphi 2\right| \right) =-2\pi
V_0\delta \left( 2\left| \sin \frac \varphi 2\right| \right) ,$ and choose $%
V_{e-h}^{SW,Q}(\varphi )=-2\pi V_0\delta (\varphi -\varphi _Q)$ (see Fig.
(2a)). By solving the one-dimensional Schr\"{o}dinger equation 
\[
\left\{ -\frac{\hbar ^2}{m^{*}R_0^2}\frac{\partial ^2}{\partial \varphi ^2}%
-2\pi V_0\delta (\varphi -\varphi _Q)-\varepsilon _j^{r.a.}\right\} \phi
_j(\varphi -\varphi _Q)=0, 
\]
we obtain one bound eigenstate with the eigenfunction and eigenenergy 
\begin{eqnarray}
\phi _0(\varphi -\varphi _Q) &=&\sqrt{k}\exp (-k\left| \varphi -\varphi
_Q\right| ),  \label{deltasw} \\
\varepsilon _0^{r.a.} &=&-\frac{\hbar ^2k^2}{m^{*}R_0^2}=-\pi V_0k,
\end{eqnarray}
respectively. Here $k\equiv \frac{\pi V_0}{\left( \frac{\hbar ^2}{m^{*}R_0^2}%
\right) }\gg 1.$ Substituting the single well wave function (\ref{deltasw})
in Eq. (\ref{tj}), in the limit of large $k$, we get the result 
\begin{eqnarray*}
t_{0,j} &\approx &4\pi V_0k\exp \left( -4\pi k\right) , \\
t_{1,j} &\approx &2\pi V_0k\exp (-2\pi k),
\end{eqnarray*}
and 
\begin{equation}
\varepsilon _0^{r.a.,p}\approx -\pi V_0k\left[ 1+4\cos \left( \frac{2\pi
\Phi }{\Phi _0}\right) \exp (-2\pi k)\right] .
\end{equation}
Note that the expression for $\varepsilon _0^{r.a.,p}$ agrees exactly with
that obtained by R\"{o}mer and Raikh (Eq. (19) in Ref. (\cite{romer})).

\subsection{harmonic potential}

In this case the electron-hole interaction potential and single well
potential are given by 
\begin{eqnarray*}
V_{e-h}\left( 2R_0\left| \sin \frac \varphi 2\right| \right) &=&\frac{%
m^{*}R_0^2}4\omega ^2\left( 2\sin \frac \varphi 2\right) ^2, \\
V_{e-h}^{SW,Q}(\varphi ) &=&\frac{m^{*}R_0^2}4\omega ^2(\varphi -\varphi
_Q)^2,
\end{eqnarray*}
respectively. A Schematic plot of these two potentials is shown in Fig.
(2b). Once again we should solve the one-dimensional Schr\"{o}dinger
equation 
\[
\left\{ -\frac{\hbar ^2}{2I}\frac{\partial ^2}{\partial \varphi ^2}+\frac{%
I\omega ^2}2(\varphi -\varphi _Q)^2-\varepsilon _j^{r.a.}\right\} \phi
_j(\varphi -\varphi _Q)=0, 
\]
where $I=\frac{m^{*}R_0^2}2$. It is nothing but an equation describing a
shifted harmonic oscillator. The eigenfunctions and eigenenergies are thus
given by 
\begin{eqnarray}
\phi _j(\varphi -\varphi _Q) &=&\frac 1{\sqrt{2^jj!\pi ^{1/2}\xi }}\exp
\left[ -\frac 12\left( \frac{\varphi -\varphi _Q}\xi \right) ^2\right] 
\nonumber \\
&&\times H_j\left( \frac{\varphi -\varphi _Q}\xi \right) ,
\label{harmonicsw} \\
\varepsilon _j^{r.a.} &=&\hbar \omega \left( j+\frac 12\right) ,\qquad
j=0,1,2,...  \nonumber
\end{eqnarray}
In Eq. (\ref{harmonicsw}), $\xi \equiv \left( \hbar \omega /2\left( \frac{%
\hbar ^2}{m^{*}R_0^2}\right) \right) ^{-1/2}\ll 1$ is the typical width of
the single well wave function and $H_j(x)$ is Hermite polynomial \cite{math}%
. By substituting Eq. (\ref{deltasw}) in Eqs. (\ref{tj}) and (\ref{miniband}%
), in the limit of small $\xi $, the final result is 
\begin{eqnarray*}
t_{0,j} &\approx &\alpha _j\hbar \omega \\
t_{1,j} &\approx &(-)^j\beta _j\exp \left( -\frac{\pi ^2}{\xi ^2}\right)
\hbar \omega ,
\end{eqnarray*}
and 
\begin{eqnarray*}
\varepsilon _j^{r.a.,p} &\approx &\hbar \omega \left[ \left( j+\frac 12%
\right) \right. \\
&&\left. -\alpha _j+(-)^{j+1}2\beta _j\exp \left( -\frac{\pi ^2}{\xi ^2}%
\right) \cos \left( \frac{2\pi \Phi }{\Phi _0}\right) \right] ,
\end{eqnarray*}
where 
\begin{eqnarray*}
\alpha _j &=&\frac{\xi ^2}{24}\frac 1{2^jj!\pi ^{1/2}}\int\limits_{-\infty
}^{+\infty }dxx^4H_j^2(x)\exp (-x^2), \\
\beta _j &=&\frac{(\pi ^2-4)}{2^{j+1}j!\xi ^2}H_j^2(\frac \pi \xi ).
\end{eqnarray*}

\subsection{Coulomb interaction potential}

Let us consider the more realistic Coulomb interaction potential which has
the form (see also Fig. (2c)) 
\begin{eqnarray*}
V_{e-h}\left( 2R_0\left| \sin \frac \varphi 2\right| \right)  &=&-\frac{e^2}{%
4\pi \varepsilon _0\varepsilon _rR_02\left| \sin \frac \varphi 2\right| }, \\
V_{e-h}^{SW,Q}(\varphi ) &=&-\frac{e^2}{4\pi \varepsilon _0\varepsilon
_rR_0\left| \varphi -\varphi _Q\right| }.
\end{eqnarray*}
We now face to solve the Schr\"{o}dinger equation for an one-dimensional
exciton, 
\[
\left\{ -\frac{\hbar ^2}{m^{*}R_0^2}\frac{\partial ^2}{\partial \varphi ^2}-%
\frac{e^2}{4\pi \varepsilon _0\varepsilon _rR_0\left| \varphi -\varphi
_Q\right| }-\varepsilon _j^{r.a.}\right\} \phi _j(\varphi -\varphi _Q)=0.
\]
The ground state solution gives a logarithmically divergent eigenenergy ($%
\varepsilon _j^{r.a.}\rightarrow -\infty )$ and the normalized eigenfunction
behaves like a $\delta $ function \cite{loudon}. In this sense, the
electron-hole pair tends to bind extremely tightly, and is prevented to
tunnel through the mean Coulomb potential barrier to induce any measurable
AB effect. This situation is also expected if one introduces a finite but
small ring width $W$. On the other hand, each excited state of
one-dimensional exciton is two-fold degenerate and classified further by the 
{\em parity} symmetry parameter $\theta $ ($\theta =0,1$ corresponds to even
and odd parity symmetry, respectively.). The detailed eigenfunctions can be
constructed from the radial wave function of a hydrogen atom with zero
angular momentum, and are given by 
\begin{eqnarray}
\phi _{j\theta }(\varphi -\varphi _Q) &=&\left\{ 
\begin{array}{ll}
\psi _j(\varphi -\varphi _Q) & \varphi >\varphi _Q \\ 
\left( -\right) ^\theta \psi _j(\varphi -\varphi _Q) & \varphi <\varphi _Q
\end{array}
\right. ,\quad j=1,2,...  \nonumber \\
\psi _j(\varphi -\varphi _Q) &=&\left( \frac 2{j^3a^3}\right) ^{1/2}\left|
\varphi -\varphi _Q\right| \exp \left( -\frac{\left| \varphi -\varphi
_Q\right| }{ja}\right)   \nonumber \\
&&\times F\left( -j+1,2,\frac{2\left| \varphi -\varphi _Q\right| }{ja}%
\right) ,  \label{coulombsw}
\end{eqnarray}
where $a\equiv 2\left( \frac{\hbar ^2}{m^{*}R_0^2}\right) /\left( \frac{e^2}{%
4\pi \varepsilon _0\varepsilon _rR_0}\right) $ $\ll 1$ and $F$ is a
hypergeometric function \cite{math}. The corresponding eigenenergies have
the form 
\[
\varepsilon _{j\theta }^{r.a.}=-\frac 1{aj^2}\frac{e^2}{8\pi \varepsilon
_0\varepsilon _rR_0},
\]
similar to the energy spectra of a hydrogen atom. It should be pointed out
that the eigenfunctions vanish at $\varphi =\varphi _Q$ since the singular
and non-integrable Coulomb potential acts as an impenetrable barrier for
both electron and hole \cite{andrews}. This effect gives an exponentially
small oscillator strength for the excited states (see Eq. (\ref{osc})) and
leads to the rapid damping of the low-lying excitonic transitions in the
linear optical susceptibility for a finite width nano-ring as shown in the
next section. The hopping integrals in Eq. (\ref{tj}) are tedious to
integrate, and in the first order of small-$a$ limit the final answer is 
\begin{eqnarray*}
t_{0,j} &\approx &\alpha _j\frac{e^2}{8\pi \varepsilon _0\varepsilon _rR_0},
\\
t_{1,j} &\approx &(-)^\theta \beta _j\exp \left( -\frac{2\pi }{ja}\right) 
\frac{e^2}{8\pi \varepsilon _0\varepsilon _rR_0},
\end{eqnarray*}
and 
\begin{eqnarray*}
\varepsilon _{j\theta }^{r.a.,p} &\approx &-\frac{e^2}{8\pi \varepsilon
_0\varepsilon _rR_0}\left[ \frac 1{aj^2}\right.  \\
&&\left. +\alpha _j+(-)^\theta 2\beta _j\exp \left( -\frac{2\pi }{ja}\right)
\cos \left( \frac{2\pi \Phi }{\Phi _0}\right) \right] ,
\end{eqnarray*}
where 
\begin{eqnarray*}
\alpha _j &=&\frac{ja}{48}\int\limits_0^{+\infty }dxx^3\left[ F\left(
-j+1,2,x\right) \right] ^2\exp (-x), \\
\beta _j &=&\frac{2\pi (\pi -2)}{j^3a^3}\left( 2\pi +ja\right) \left[
F\left( -j+1,2,2\pi /ja\right) \right] ^2.
\end{eqnarray*}

At the end of this section, we would like to emphasize the following
features:

(i) For all the cases we have considered, both for ground state and excited
states the hopping integral $t_{1,j}$ is typically exponentially small
compared with $\varepsilon _j^{r.a.}$, and the exponent factor is determined
by the ratio of electron-hole interaction potential and the characteristic
energy of rotation mode. This prediction is in apparent contradiction to the
recent claim of R\"{o}mer and Raikh \cite{romer}, which predicted that the
ABE oscillations are found to be much more easily in the case of excited
states, and are not exponentially small. This discrepancy might arise from
the definition of ''excited states'' of excitons. Recall that there is only
one bound state for electron-hole pair with short-ranged interaction
potential (at least with $\delta $ potential). Therefore, the ''excited
states'' said by R\"{o}mer and Raikh is more appropriate to be regarded as
''excited states'' of {\em free} electrons or holes instead of excitons.

(ii) In the case of the Coulomb interaction potential, there is no ABE for
the ground excitonic state due to the intrinsic divergence of the ground
state energy of one-dimensional excitons \cite{loudon}.

(iii) On the other hand, in order to observe the AB effect in excited
states, the single well eigenenergies $\varepsilon _j^{r.a.}$ should be
tuned closely to the potential maximum $V_{e-h}\left( \pi \right) $. Because
of the particular properties of the energy spectra in the presence of
Coulomb interaction, {\em i.e.}, $\varepsilon _{j\theta }^{r.a.}\propto
-1/j^2$, the Coulomb interaction potential is appropriate to observe the AB
effect in excited states compared with other potentials. For example, for a
narrow-width nano-ring with $R_0=20$ {\rm nm}, by setting $\varepsilon
_{j\theta }^{r.a.}\sim -\frac{e^2}{8\pi \varepsilon _0\varepsilon _rR_0},$ a
rough estimation gives $j\sim 1$, which agrees well with the numerical
results for light-hole-excitons in InAs nano-rings as shown below.

\section{NUMERICAL\ RESULTS AND DISCUSSIONS}

In order to understand the influence of a finite ring width on the AB
effect, let us now discuss the numerical results for a {\em realistic}
self-assembled semiconducting InAs nano-ring. In the following we restrict
ourselves in the subspace $L=0$. We have taken the parameters $%
m_e^{*}=0.067m_e$, the effective mass of the light hole $m_{lh}^{*}=0.099m_e$%
, the effective mass of the heavy hole $m_{hh}^{*}=0.335m_e$ ($m_e$ is the
bare mass of a single electron) and $\varepsilon _r=12.4$, which are
appropriate to InAs material \cite{lorke00,emperador,hu}. The electron and
hole are considered to be confined in a confinement potential with the same
strength, {\em i.e.}, $m_e^{*}\omega _e^2=m_h^{*}\omega _h^2$. The ring
radius $R_0$ and characteristic frequency of the radial confinement $\hbar
\omega _e$ are chosen to be $20$ {\rm nm} and $14$ {\rm meV }($\hbar \omega
_e=14$ {\rm meV} corresponds to $W=10$ {\rm nm}), respectively, simulating a
recent experimental results $\Delta E_m\sim 5$ {\rm meV} and $\Delta E_n\sim
20-25$ {\rm meV }in InAs nano-rings \cite{pettersson}, $\Delta E_m$ is the
energy level spacing between the single electron states with different
orbital angular momentum $m$ and the same radial quantum number $n$, while $%
\Delta E_n$ corresponds to the energy spacing with different radial quantum
number $n$ and the same angular momentum $m$. In the calculations, we use
effective atomic units in which the length unit $a_B^{*}$ is a factor $\frac{%
\varepsilon _r}\mu $ times the Bohr radius $a_B$, and the energy is given in
effective hartrees, $H_a^{*}=(\mu /\varepsilon _r^2)\times 1$ hartree (For
the heavy-hole-exciton, for example, the length and energy units then scale
to $a_B^{*}=11.8$ {\rm nm} and $H_a^{*}=10.0$ {\rm meV}). For $R_0=20$ {\rm %
nm, }the universal flux quantum $\Phi _0$ corresponds to the magnetic field $%
B\approx 4.1$ ${\rm T.}$

By probing of the structure of the numerical excitonic wave functions with
the use of the conditional probability distribution (CPD) and
pair-correlation function (PCF) \cite{yann00a,yann00b}, we first study the
hopping integral $t_{1,j}$ (or the tunneling possibility) which plays the
most essential role in the AB effect. Denoting the numerical wave function
of excitons by $\Psi \left( {\bf r}_e,{\bf r}_h\right) $ (see Eq. (\ref
{numwvf}), which is the summation of the product of the centre-of-mass and
relative wave functions), we define the usual PCF as 
\begin{equation}
G(\upsilon )=2\pi \int \int \delta \left( {\bf r}_e-{\bf r}_h-{\bf v}\right)
\left| \Psi \left( {\bf r}_e,{\bf r}_h\right) \right| ^2d{\bf r}_ed{\bf r}_h,
\label{pcf}
\end{equation}
and the CPD for finding the electron at ${\bf v}$ given that the hole is at $%
{\bf r}_h={\bf v}_0$ as 
\begin{equation}
{\cal P}({\bf v\mid r}_h={\bf v}_0)=\frac{\left| \Psi \left( {\bf v},{\bf r}%
_h={\bf v}_0\right) \right| ^2}{\int d{\bf r}_e\left| \Psi \left( {\bf r}_e,%
{\bf r}_h={\bf v}_0\right) \right| ^2}.  \label{cpd}
\end{equation}
Note that for the exact $\Psi $ in the case of a circularly symmetric
confinement, the PCF turn out to be circularly symmetric.

With the above, we solved for heavy-hole-exciton energy spectra and wave
functions for $W=10$ {\rm nm}. The selected PCF's and CPD's are displayed in
Fig. 3. For the ground state in Fig. (3a), the CPD exhibits a highly
localized electron density around the hole, namely, the electron and hole
tends to bind tightly with each other. The above picture is also reflected
in the PCF, which shows a rapid delay as the distance between electron and
hole increases. This behaviour agrees well with the rigorous analysis
present earlier in Sec IIIC. In contrast to the ground state, in Figs. (3b)
and (3c) the CPD's for the fifth and sixth excited states exhibit a
well-developed local maximum probability for finding the electron around the
diametrically opposite point $(-8,0)$, suggesting that the electron and hole
have a possibility to tunnel through the Coulomb potential barrier in the
opposite directions. Observe also that the PCF's in Figs. (3b) and (3c)
reveal a clear local maximum at $\upsilon \sim 35$ {\rm nm}. Moreover, in
the case of the eighth and ninth excited states (see Fig. 4), one can even
identify a larger hopping integral. The remarkable emergence of the hopping
integral for low-lying excited excitonic states provides us a possibility
for observing a non-vanishing AB effect.

Figs. (5a) and (5b) display the low-lying energy levels as a function of the
magnetic field for heavy-hole-exciton and light-hole-exciton. As can be seen
immediately, the most obvious feature is the AB oscillations of some
low-lying energy levels (indicated by an arrow) with a period $B\approx 4.1$ 
${\rm T}$ or a flux quantum $\Phi _0$, as expected. However, the AB
oscillations are not very close to sinusoidal as illustrated by Eq. (\ref
{miniband}), due to the finite width of nano-rings. The typical amplitude of
those oscillations is about $3$ {\rm meV}, which is much larger than the
extremely narrow luminescence line-width\ in a {\em single} InAs nano-ring 
\cite{warburton}, and in view of this it might be sensitive to be detected
by present optical emission techniques. Comparing Figs. (5a) and (5b), it is
seen that the energy levels of light-hole-exciton show a relatively
pronounced AB effect. Even the energy level of the first excite state of
light-hole-exciton tends to oscillate with magnetic field. This is due the
comparable effective mass between electron and light-hole, which suppresses
the electron-hole binding ({\em i.e.}, the much higher ground state energy
of light-hole-exciton compared with that of heavy-hole-exciton) and thus
enhances the AB effect.

Another noticeable features shown in Figs. (5a) and (5b) include: (i) There
are many anticrossings in the energy levels which is caused by level
repulsions. (ii) Contrary to conventional quantum dots, when the magnetic
field is increased it seems unlikely to form any Landau levels. (iii) In the
high magnetic field, the energy level of the ground state and first excited
state shows a slight blue shift, due to the diamagnetic effect for each
charge carrier which pushes all the relative energies upwards. This behavior
is qualitatively similar to the excitons in quantum dots. Note that those
blue shifts are observed in a recent experiment \cite{hakan}.

The behavior of the oscillator strengths for heavy-hole-exciton and
light-hole-exciton as a function of the magnetic field is present in Figs.
(6a)-(6d). It is readily seen that periodic oscillations take place in the
oscillator strengths for the selected low-lying excited states. Though those
oscillations are not regular in shape and not close to the expected
sinusoidal, they are still supposed to be caused by AB phenomena. In
contrast, for the ground state, the oscillator strength increase
monotonously with the magnetic field. The result is in line with the blue
shifts of the ground-state energies, for example, the increasing magnetic
field increases the confinement, decreases the separation between electron
and hole and in turn yields the enhancement of oscillator strengths.
Furthermore, the overall magnitude of the oscillator strength of the ground
state is approximately one order larger that of low-lying excited states, in
good agreement with the analysis present earlier that the oscillator
strengths for excited states are exponentially small for ideal
one-dimensional nano-rings.

To support the experimental relevance of our results and better understand
the periodic, Aharonov-Bohm-type oscillation in energy spectra, the {%
imaginary part of linear optical susceptibility is plotted as a function of
frequency }$\omega $ for different magnetic fields in Figs. (7a)-(7d), where
a broadening parameter $\Gamma =0.5$ {\rm meV} is used. Those curves
represent all the possible transitions of excitonic states which would be
measurable via photoluminescence excitation measurements (PLE). As expected,
the periodic oscillations at some low-lying energy levels ($\omega \sim 30$ 
{\rm meV) }are well-reflected. However, the amplitude of those oscillations
is very small compared with that of the fundamental transition because of
the weak oscillator strengths of low-lying excited states. In fact, it is an
indication of the delicate nature of the AB effect, which suggests that a
highly sensitive technique is needed in experiment to access these coherent
AB oscillations.

It is important to point out that our results mentioned above are in
apparent contradiction to the prediction given by Song and Ulloa in their
recent works, in which they claimed that the excitons in nano-rings behave
to a great extent as those in quantum dots of similar dimensions and the AB
oscillation of exciton characteristics predicted for one-dimensional rings
are found to not be present in finite-width systems \cite{song00}. Here we
indeed observe the ABE of some of low-lying exciton states, which is
suggested by the CPD's and PCF's, illustrated by the energy spectra and
oscillator strengths and finally confirmed by the {imaginary part of linear
optical susceptibility.} Therefore, we believe that the finite but small
width of the nano-rings will not suppress the ABE so greatly. Since the
parameters used by Song and Ulloa are quite similar to ours, we suggest that
the discrepancy might originate from the following reasons:

(i) In their paper, the negative Coulomb interaction term has been simply
treated as a perturbation within a parabolic basis. However, just as the
Authors mentioned, when the characteristic scale of systems is beyond the
effective Bohr radius, the known poor-convergence of the parabolic basis
might lead to unreliable results, especially for the excited states.

(ii) For the radius of ring in the Fig. 4 in Ref. (\cite{song00}), $R_0=24$ 
{\rm nm}, the periodicity of AB oscillation is expected to be given by a
period $\Delta B\approx 2.9$ {\rm T}. However, the Authors only show the
results for different magnetic fields increased on steps of $5.0$ {\rm T}.
With so large interval, one might miss the AB oscillations completely..

We next turn out to investigate the size effect by tunning the ring width.
As an illustration, the low-lying energy levels of the heavy-hole exciton
for a narrow ring width $W=8$ {\rm nm}, shown as curve (a) in Fig. 8, is
compared to the wide ring width case with $W=14$, $20$ and $30$ {\rm nm}
(shown in Fig. 8 as curves (b), (c) and (d), respectively). It is readily
seen that as the ring width increases, the AB oscillation pattern is
gradually destroyed. The disappearance of AB effect arises from the
destruction of the non-simply-connected geometry of nano-rings, since the
increasing width lowers the soft confinement potential barrier at the ring
center and in turn yields a high possibility for carriers to reside. For $%
W=30$ {\rm nm}, which is comparable to the ring diameter $d=2R_0=40$ {\rm nm}%
, the main characteristic of the energy spectra resembles that of quantum
dots \cite{halonen92,wojs,hawprb99}. In Fig. 9, the corresponding {imaginary
part of linear optical susceptibility is plotted }as a function of
frequency. Comparison of the case of $W=8$ {\rm nm} in Fig. (9a) with the
wide width limit (Fig. (9d)) is instructive, since it exhibits a distinctive
difference of the excitonic optical properties between nano-rings and
quantum dots. Unlike the conventional quantum dots, in which the low-lying
exciton state transitions have the same amplitudes and are nearly equally
distributed (a reflection of excitations involving the exciton ground state
and various center-of-mass replicas without altering the ground state of the
relative coordinate), the low-lying transitions of nano-rings show a rapid
dampen with frequency and theirs positions are not periodic. This difference
is a refection of the anisotropic confinement of nano-rings \cite{song00}:
Since the exciton is confined in a quasi-one-dimensional system, its
center-of-mass degree of freedom is greatly suppressed and its relative
motion becomes dominant ({\em i.e.} the general properties of
one-dimensional exciton is expected), thus resulting in the destruction of
the regular patterns observed in quantum dots. It is interesting to note
that this distinctive difference is indeed observed by a recent experiment
given by Pettersson {\em et al. }\cite{pettersson,hu00a,hu00b}.

\section{CONCLUSION}

In conclusion, we have studied the magnetic field effect on excitons in an
InAs nano-ring based on a simple model Hamiltonian. By numerical
diagonalization, we calculate the low-lying energy levels, oscillator
strengths, and the corresponding linear optical susceptibility. A periodic,
Aharonov-Bohm-type oscillation is clearly revealed in some low-lying energy
levels and linear optical susceptibility curves for a {\em realistic}
self-assembled semiconducting InAs nano-ring. This AB effect is well
interpreted with the rigorous analysis for narrow-width nano-rings.

In a recent work, Warburton {\em et al.} presented a beautiful experiment of
optical emission in a {\em single} charge-tunable nano-ring \cite{warburton}%
. They studied the role of multiply-charged exciton complexes with no
applied magnetic field and found a shell structure in energy similar to that
of quantum dots \cite{bayer,hawprl00}. Therefore, encouraged by the rapid
developed nano-techniques for detection, {\em i.e.}, the achievement of
extremely narrow and temperature insensitive luminescence lines from a
single InAs nano-ring in GaAs, we hope that our predictions of the ABE
effects can be confronted in experiments in the future.

\begin{center}
{\bf ACKNOWLEDGMENTS}
\end{center}

We would like to thank Dr. H. Pettersson for helpful discussion and Dr. A.
V. Chaplik for sending a copy of the paper (Ref. \cite{chaplik}). This
research is financially supported by the NSF-China (Grant No.19974019) and
China's ''973'' program.

\begin{center}
{\bf Figures Captions}
\end{center}

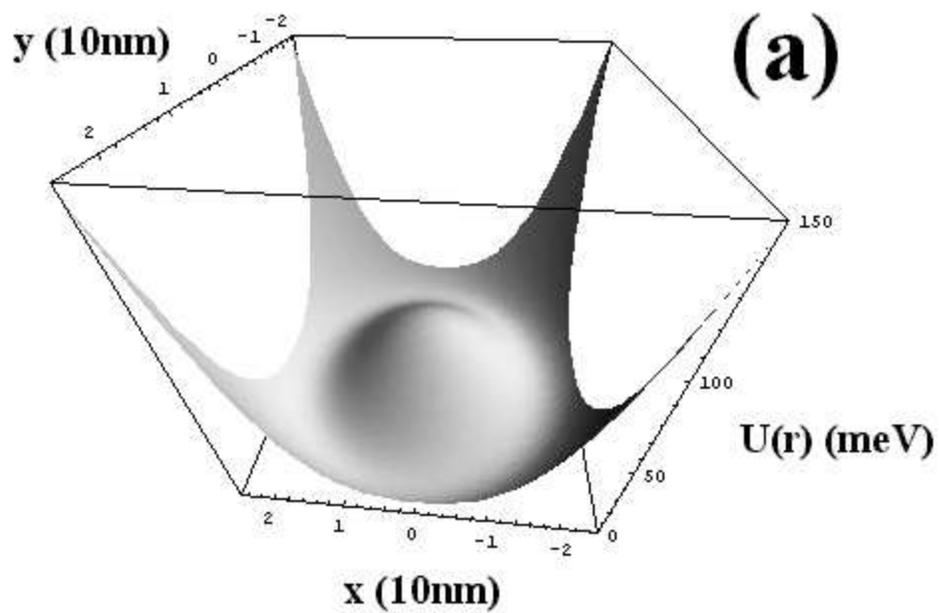
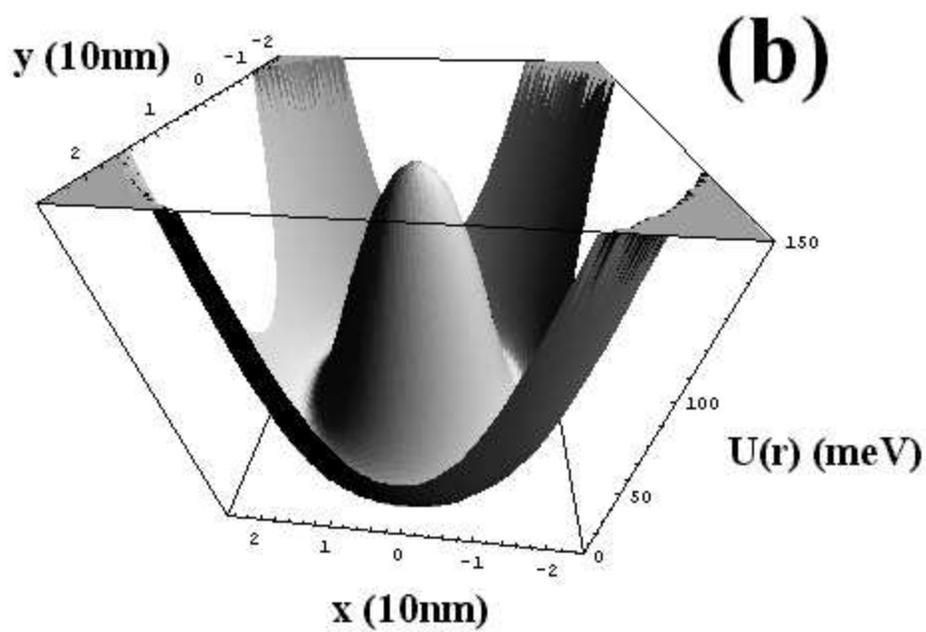
Fig.1. Schematic geometry of InAs nano-rings with radius $R_0=20$ {\rm nm}
and width $W=10$ {\rm nm} (a) or $7$ {\rm nm} (b) for electrons. $%
m_e^{*}=0.067m_e.\newline
$

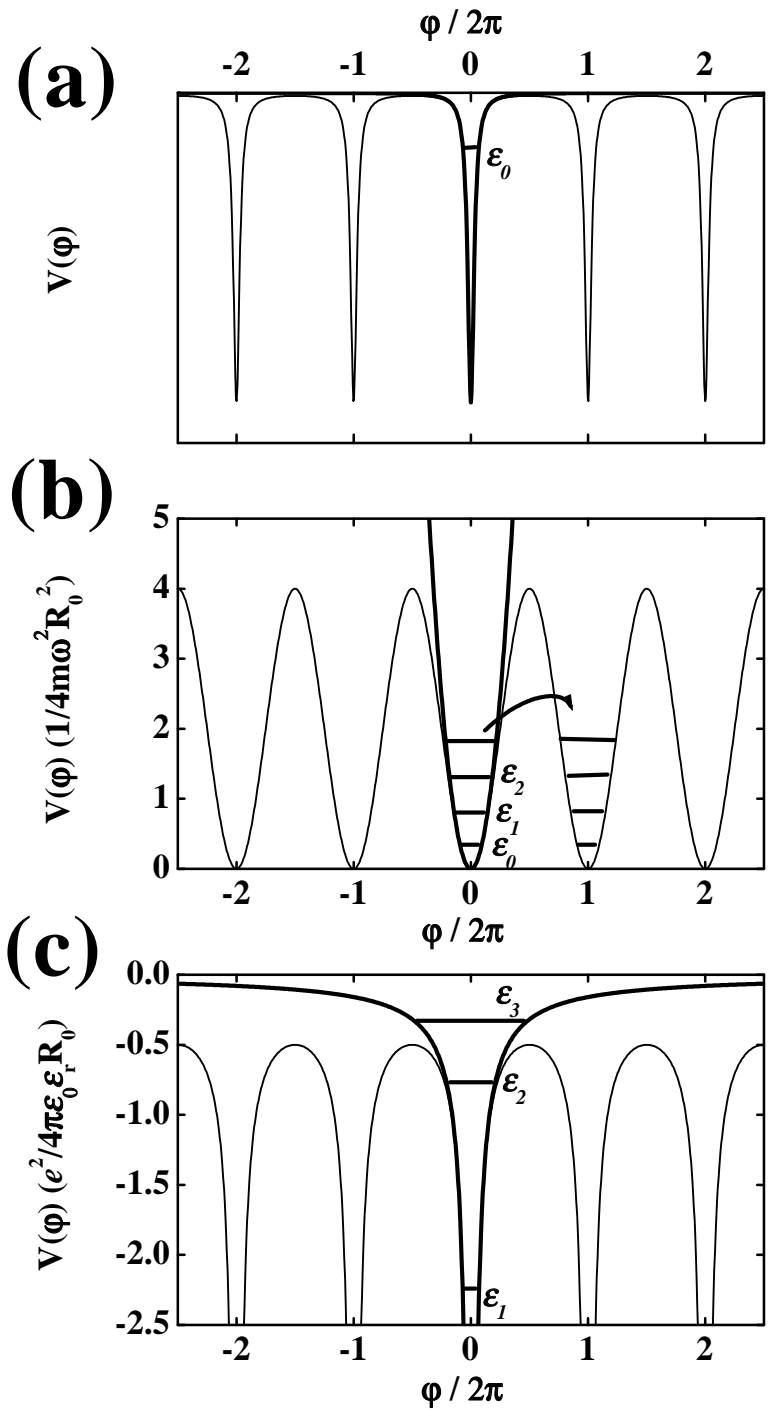
Fig.2. Scheme of different interaction potentials $V_{e-h}(\varphi )$ as a
function of the relative azimuthal coordinate $\varphi =\varphi _e-\varphi
_h $. (a) Short-ranged form $-2\pi V_0\delta (2\sin \frac \varphi 2)$, (b)
parabolic form $\frac{\mu \omega ^2\rho ^2}4\left( 2\sin \frac \varphi 2%
\right) ^2$ and (c) Coulomb form $-\frac{e^2}{4\pi \varepsilon _0\varepsilon
_r\rho \left| 2\sin \frac \varphi 2\right| }.$ The potential of a given
single well with minimum at $\varphi =0$, $V_{e-h}^{SW,0}(\varphi )$ is
delineated by the thick solid line. Some low-lying energy levels for each
well are also indicated.\newline

Fig.3. The conditional probability distributions (CPD's) and
pair-correlation functions (PCF's) shown, respectively, on the left and
right of each of the subplots [labeled by (a)-(c)], for heavy-hole exciton
in an InAs nano-ring with radius $R_0=20$ {\rm nm} and width $W=10$ {\rm nm}%
. Each CPD is expressed in a logarithmic intensity scale with the range from 
$0.01$ (black) to $1.0$ (white). The heavy-hole is fixed at ${\bf v}%
_0=(20,0) $. (a) The ground state, (b) the fifth and (c) sixth excited
states.\newline

Fig.4. The conditional probability distributions (CPD's) for the eighth
(upper panel) and ninth (lower panel) excited states. Other parameters are
the same as in Fig. 3.\newline

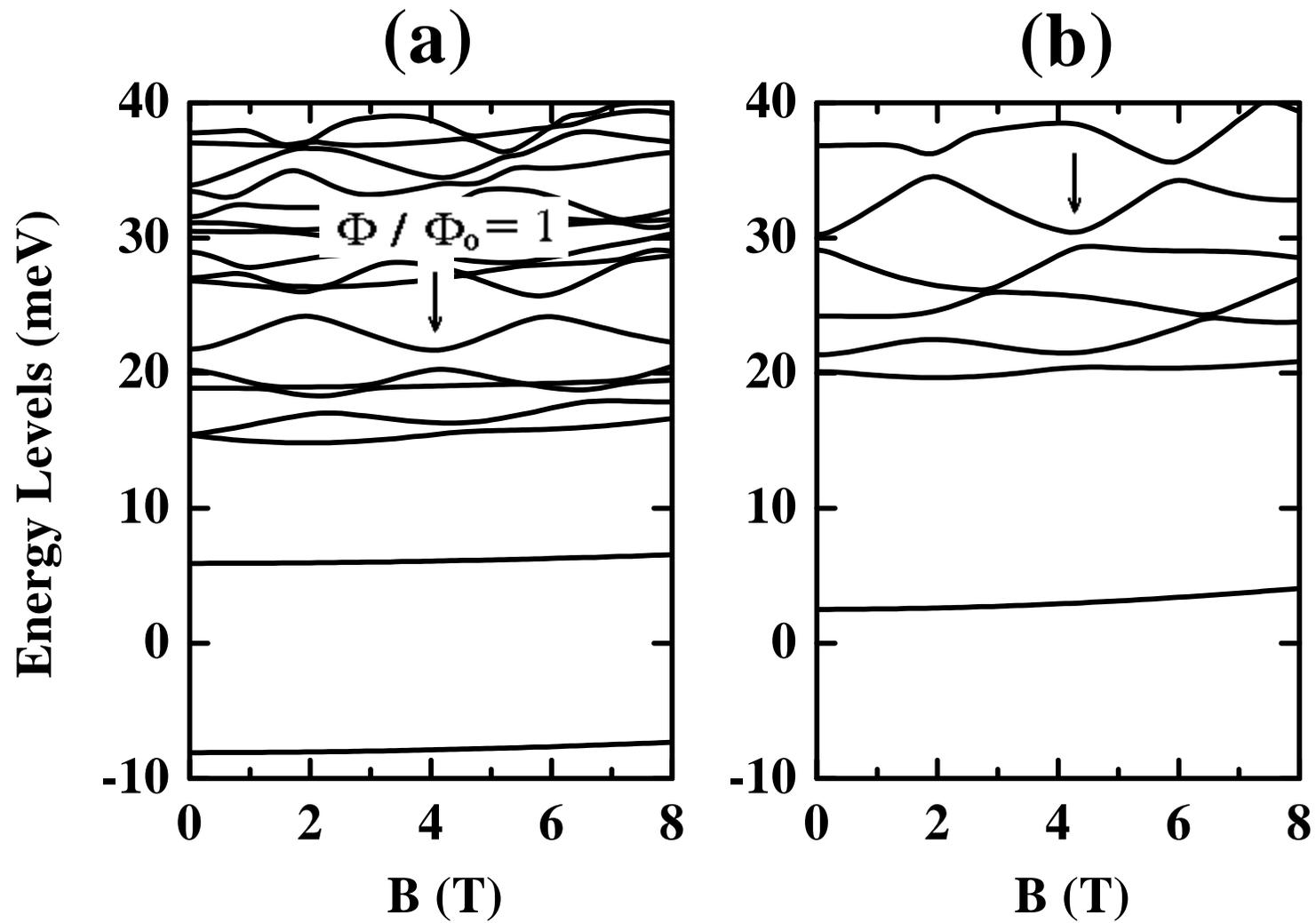
Fig.5. The low-lying energy levels of InAs nano-rings with radius $R_0=20$ 
{\rm nm} and width $W=10$ {\rm nm }as a function of the magnetic field for
heavy-hole (a) and light-hole (b) excitons. As indicated by the arrow, a
periodic, Aharonov-Bohm-type oscillation in some low-lying energy levels
takes place.\newline

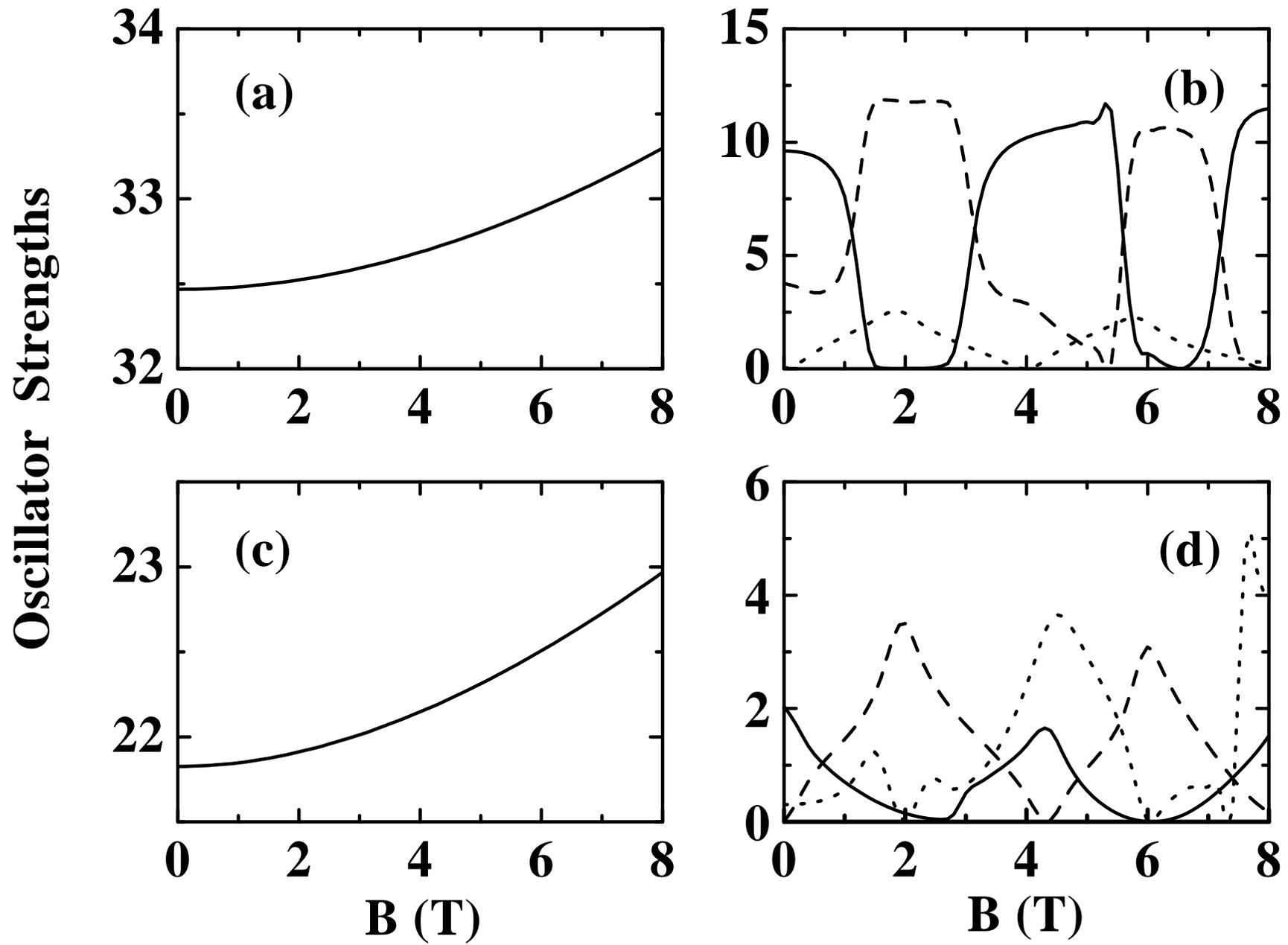
Fig.6. The oscillator strengths of InAs nano-rings as a function of the
magnetic field for heavy-hole (upper panel, (a) and (b)) and light-hole
(lower panel, (c) and (d)) excitons. (a) (c) for ground state and (b) (d)
for excited states (the solid, dashed and dotted lines correspond to the
fourth, fifth and sixth excited states).\newline

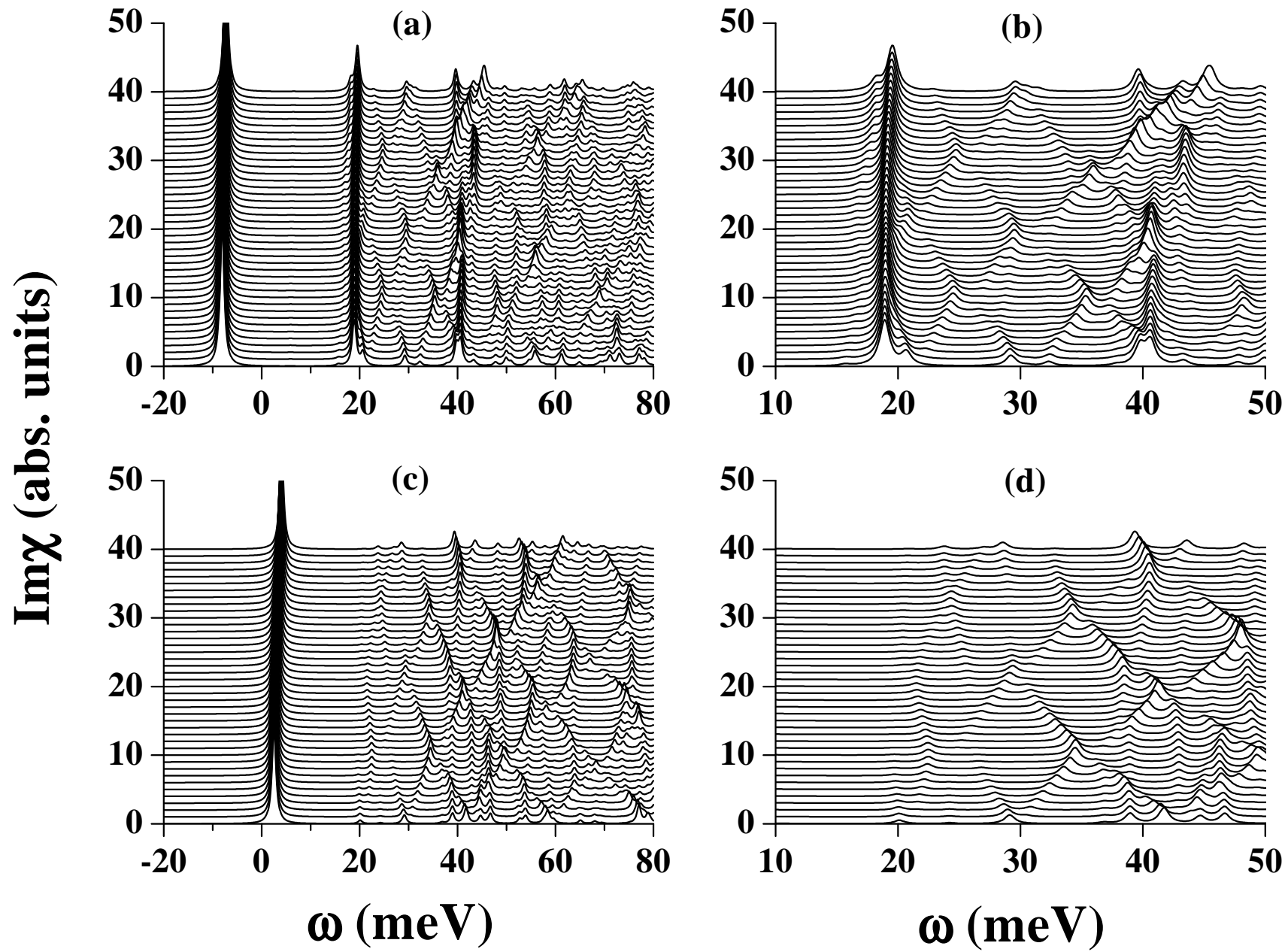
{Fig.7. Imaginary part of linear optical susceptibility }as a function of
frequency for different magnetic fields (in each subplot, from bottom to
top, $B=0.0$ $\sim 8.0$ ${\rm T}$ is increased on steps of $0.2$ {\rm T}).
(a),(b) for heavy-hole and (c),(d) for light-hole excitons. (b) and (d) are
the enlarged versions of (a) and (c). Ring radius $R_0=20$ {\rm nm} and
width $W=10$ {\rm nm. }The semiconducting band gap $E_g$ is taken to be
zero. Note that for the sake of clarity curves have been upshifted with a
constant.\newline

Fig.8. The low-lying energy levels of InAs nano-rings as a function of the
magnetic field for heavy-hole exciton with different ring widths: (a) $W=$ $%
8 $ {\rm nm}, (b) $14$ {\rm nm}, (c) $20$ {\rm nm} and (d) $30$ {\rm nm. }%
Other parameters are the same as in Fig. 5.\newline

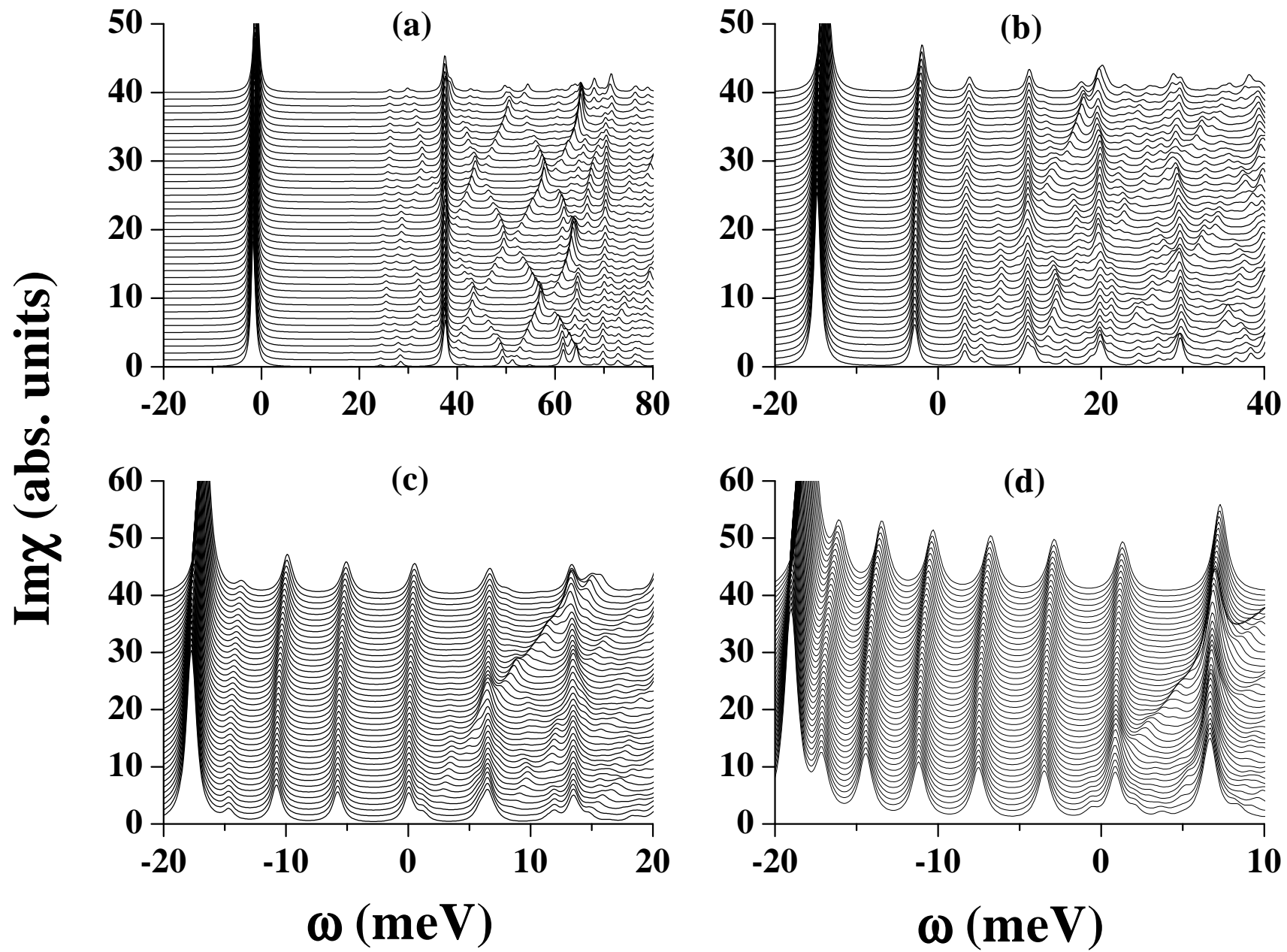
Fig.9. {Imaginary part of linear optical susceptibility of heavy-hole
exciton }as a function of frequency for different magnetic fields and
different ring widths: (a) $W=$ $8$ {\rm nm}, (b) $14$ {\rm nm}, (c) $20$ 
{\rm nm} and (d) $30$ {\rm nm.}


\begin{references}
\bibitem{note}  Upon completion of this work, we finally obtained the Ref. (%
\cite{chaplik}), in which the main result is similar to the one in Sec. IIIC.
\end{references}
\end{document}